\documentclass[a4paper,11pt]{article}
\usepackage{pos}

\title{NLO event generation for LHC neutrinos and application to 
flux measurements at FASER}
\ShortTitle{NLO event generation for LHC neutrinos}

\author*[a,b]{Peter Krack}

\affiliation[a]{Department of Physics and Astronomy, Vrije Universiteit, \\
NL-1081 HV Amsterdam, The Netherlands}
\affiliation[b]{Nikhef Theory Group, \\ Science Park 105, 1098 XG Amsterdam, The Netherlands}

\emailAdd{peter.krack@nikhef.nl}

\abstract{The LHC generates an intense beam of high-energy neutrinos in the forward 
direction, whose scientific potential has been left unexploited for many years.
The FASER and SND@LHC experiments, operating since 2023, have recently measured LHC neutrinos for the first time.
In this contribution we discuss how to produce accurate predictions, including NLO QCD corrections and modern parton shower algorithms, for present and future LHC neutrino experiments, including those at the proposed 
Forward Physics Facility (FPF).
To this end, the energy and rapidity
distribution of the LHC neutrinos is encoded in a LHAPDF grid interfaced to
the neutrino DIS event generator in the POWHEG-BOX-RES framework.
This Monte Carlo tool enables the modelling of differential distributions that are
sensitive to hadronic final states, initial- and final-state radiation, and realistic acceptance and selection cuts. 
As a first application, we deploy this event generator to compute fast-interpolation grids and carry out a first determination of the LHC forward neutrino fluxes directly from FASER data using the   NNPDF fitting methodology.}

\FullConference{XXXII International Workshop on Deep Inelastic Scattering and 
Related Subjects (DIS2025)\\
24-28 March, 2025\\
Cape Town, South Africa\\}

\begin{document}
\maketitle

\section{Introduction}
Even though the LHC produces high-energy neutrinos of all three flavours, these 
neutrinos have for a long time been ignored, and escaped undetected in the far-forward region.
The first experiments to explore the LHC far-forward region and to observe the 
neutrinos are FASER \cite{FASER:2023zcr} and SND@LHC \cite{SNDLHC:2023pun}.
In particular, the FASER~\cite{FASER:2024ref} experiment has measured the neutrino interaction 
cross-section at TeV energies, the highest energy neutrinos produced at laboratory 
experiments so far.

Understanding these experiments require an accurate modelling of the neutrino-hadron 
interaction: the deep-inelastic scattering process~\cite{Candido:2023utz}. 
Calculations up to third 
order in perturbative QCD are available for the inclusive cross-section and 
the double-differential distributions in $x$ and $Q^2$~\cite{NNPDF:2024nan}. 
In order to study exclusive 
quantities with generic final-state cuts, a Monte Carlo event generator matched to a parton shower is required.
Most neutrino scattering experiments, including those from the LHC, have been relying on the leading-order
event generator \textsc{GENIE} \cite{Andreopoulos:2009rq}, which uses 
phenomenological Bodek-Yang model \cite{Yang:1998zb} and  the \textsc{Pythia6} 
parton shower and hadronisation algorithm \cite{Sjostrand:2000wi}.
For inclusive quantities, \textsc{GENIE} can be made NLO-aware using the HEDIS 
module~\cite{Garcia:2020jwr}, but does not allow for the incorporation of 
acceptance cuts or the matching to a PS generator.
At next-to leading order, Monte Carlo event generators using the POWHEG method
recently became available \cite{Buonocore:2024pdv,Banfi:2023mhz,vanBeekveld:2024ziz,FerrarioRavasio:2024kem}, 
which can be interfaced to the \textsc{Pythia8} general-purpose Monte Carlo event generator~\cite{Bierlich:2022pfr}.
In addition to the modelling of collider neutrino scattering, POWHEG is also able to simulate neutral-current DIS with LHC muons, whose event rates are actually much higher than those of neutrino-initiated reactions~\cite{Francener:2025pnr}.

Assuming that the neutrino-hadron interactions at TeV energies can be modelled within perturbative
QCD~\cite{Cruz-Martinez:2023sdv}, the neutrino flux can be extracted in a theory-agnostic, data-driven 
approach using machine learning (ML) methods.
In this case, the extraction of the neutrino flux is formally equivalent to the extraction
of the parton distribution functions (PDFs) of the proton.
The determination of the neutrino flux provides constraints on light- and
heavy-meson production at the LHC.

In this contribution, a phenomenological analysis of DIS at NLO is presented that 
was published in \cite{vanBeekveld:2024ziz} and recent progress on the extraction 
of the neutrino  flux from measurements at far-forward experiments is that has since 
been published in \cite{John:2025qlm} are briefly summarised.
%
%%%%%%%%%%%%%%%%%%%%%%

\section{NLO event generation at LHC far-forward experiments}
\subsection{Implementation}
The neutrino flux $\displaystyle \frac{dN_{\nu_i} (E_\nu)}{dE_\nu}$ acts in the 
generation of events like a parton distribution function for the neutrino.
Thanks to this behaviour, it can be cast into a LHAPDF grid, facilitating the 
interpolation and interface with the generator in the POWHEG-BOX-RES framework
\cite{Banfi:2023mhz}.
This can be achieved by defining the "neutrino momentum fraction" 
$\displaystyle x_\nu = \frac{2 E_\nu}{\sqrt{s_{\mathrm{pp}}}}$ with 
$0\le x_\nu \le 1$ and  where $E_\nu $ is the energy of the neutrino produced at 
the LHC in a proton-proton collision with a centre-of-mass energy of 
$\sqrt{s_{\mathrm{pp}}}$.
This yields a scale-independent "PDF set" for the three neutrino flavours 
$\displaystyle f_{\nu_i} = \frac{2}{\sqrt{s_{\mathrm{pp}}}} 
\frac{dN_{\nu_i} (E_\nu)}{dE_\nu} $ with $i=e,\mu,\tau$.

\subsection{Comparison with GENIE}
The result from simulation using \textsc{POWHEG+Pythia8} at LO and
NLO compared to GENIE, with FASER acceptance cuts and a fixed neutrino energy of
$E_\nu = 1~\mathrm{GeV}$
The HEDIS tune in GENIE was used, with the HERAPDF1.5 NLO PDF set 
\cite{Cooper-Sarkar:2010yul}.
Differences between the two tools are expected, since the hard scattering matrix
element, PDFs parton shower, and other input setting differ. 
Depending on the observable, differences up to 20\% can be observed.
\begin{figure}[!htb]
 \includegraphics[width=0.48\linewidth]{"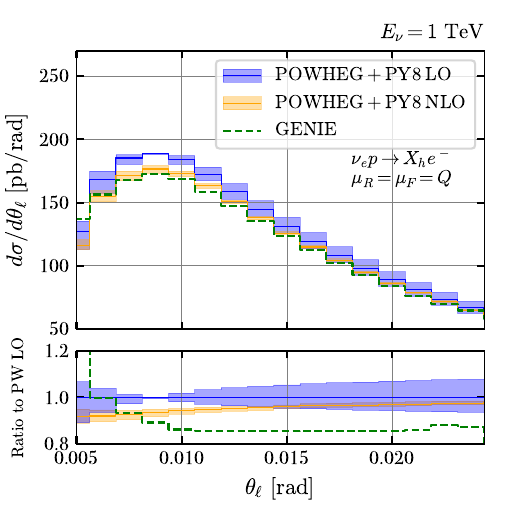"}
 \includegraphics[width=0.48\linewidth]{"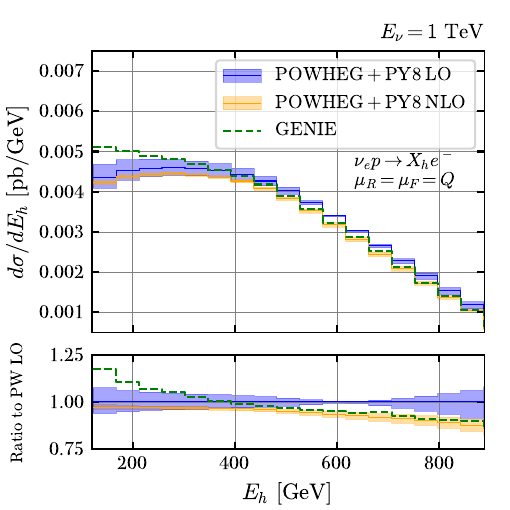"}
 \caption{\label{fig:genie} \textsc{POWHEG+Pythia8} at LO and NLO compared to 
 GENIE. The distribution for the scattering angle of the final state lepton 
 shows the most drastic disagreement, while  the energy of the hadronic state
 agrees between the two Monte Carlo tools.}
\end{figure}

\subsection{Modelling of high-energy neutrino DIS}
The impact of NLO QCD corrections, the variation of parton shower algorithm and 
tune can be studied. 
The following predictions are for the FASER$\nu$ detector, assuming cuts on the
energy of the final state lepton $E_\ell > 100~\mathrm{GeV}$, the energy of the
hadronic system $E_h > 100~\mathrm{GeV}$ and the scattering angle of the final
state lepton $\mathrm{tan}\,\theta < 0.025$.
The predictions are computed using the NNPDF4.0 NNLO PDF set and assuming isospin
symmetry to model the nuclear target.

\subsubsection{Impact of NLO QCD corrections}
\begin{figure}
 \centering
 \includegraphics[width=0.48\linewidth]{"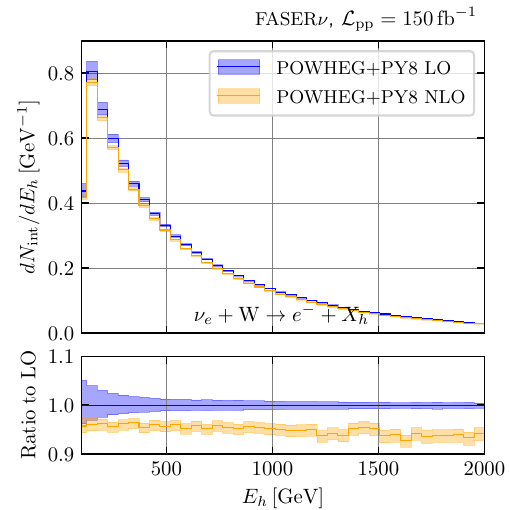"}
 \caption{\label{fig:NLOQCD} Differential distribution of the energy of the
  hadronic system, comparing the NLO and LO result.}
\end{figure}

In figure \ref{fig:NLOQCD} the differential distribution for the energy of the 
hadronic system can be found, computed at LO and NLO.
The uncertainties are determined from seven-point scale variation.
The NLO corrections can have an impact of a few percent up to 10\% in this 
case. 

\subsubsection{Impact of the parton shower algorithm}

\begin{figure}
 \centering
 \includegraphics[width=0.48\linewidth]{"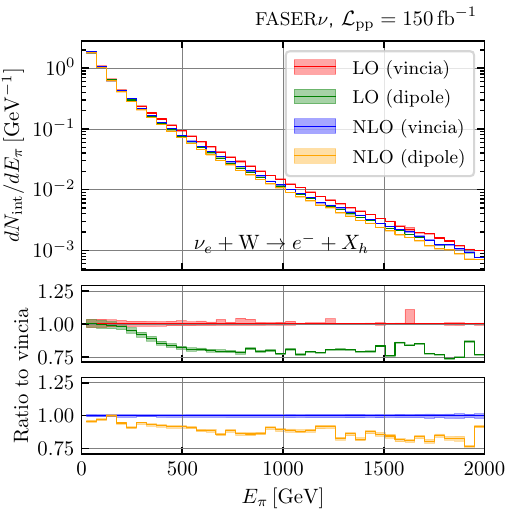"}
 \caption{\label{fig:PS} Differential distribution of the energy of the 
 leading pion for the \textsc{Pythia8} dipole and the \textsc{Vincia} 
 shower.}
\end{figure}

The choice of parton shower algorithm affects exclusive quantities that 
depend on the hadronic final state. 
Inclusive quantities like $E_\ell$, $\theta_\ell$, $Q^2$ or $E_h$ 
are largely insensitive to the choice of parton shower.
Figure \ref{fig:PS} shows the distribution for the energy of the leading pion, 
comparing the result of the \textsc{Pythia8} dipole shower with 
the \textsc{Vincia} shower.
All other settings are kept the same. 
The difference becomes smaller when showering NLO events, since the hardest
emission is already included in the matrix element.

\subsubsection{Impact of soft QCD}
\begin{figure}
 \centering
 \includegraphics[width=0.48\linewidth]{"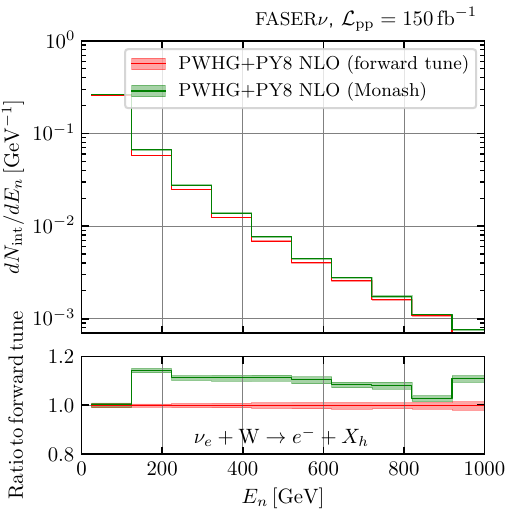"}
 \caption{\label{fig:softQCD} Differential distribution for the energy of the 
 leading neutron. The default Monash 2013 tune is compared to the LHCf tune.}
\end{figure}

Parameters that control soft and non-perturbative QCD in \textsc{Pythia8} are
set by the tune adopted by the shower algorithm. 
The effects that are modelled by the tune can further include hadronisation, 
multi-parton interactions and underlying event.
In figure \ref{fig:softQCD} the default Monash 2013 tune is compared to the 
tune from \cite{Fieg:2023kld} for forward physics to data from the LHCf 
experiment \cite{LHCf:2015nel,LHCf:2020hjf,LHCf:2015rcj,LHCf:2017fnw} for 
the energy of the leading neutron.
Especially parameters controlling the beam remnant hadronisation have be tuned.
Even though the LHCf tune is not dedicated to neutrino DIS, it can provide an 
estimate of how the predictions are affected.
Overall, a difference of around 10\% is found for the distribution of $E_n$. 
Similar results have been obtained for other exclusive quantities, such as 
$E_\pi$ and $E_K$, while inclusive quantities remained mostly unaffected by
variations of the tune.

%%%%%%%%%%%%%%%%%%%%%%

\section{Extraction of the forward neutrino flux}
The problem of extracting the neutrino flux is equivalent to extracting a PDF.
Therefore, an approach following the NNPDF methodology 
\cite{NNPDF:2014otw,NNPDF:2017mvq,NNPDF:2021njg} can be taken.
For the implementation of the FASER data, the neutrino DIS event generator 
can be used to compute the necessary fast-interface (FK-tables)~\cite{Barontini:2023vmr,Ball:2010de}.

\subsection{Methodology}
The neutrino flux is parametrised using a small neural network and a 
preprocessing function.
\begin{equation}
 f_{\nu_i}(x_\nu) = x^\alpha (1-x)^\beta \,\mathrm{NN}(x_\nu)
\end{equation}
The preprocessing function guarantees the behaviour at low- and high-$x_\nu$.
The fit of the neutrino flux is performed via a $\chi^2$ minimisation.
Uncertainties on the neutrino flux can be determined using the Monte Carlo
replica method.

\subsection{Results}
\begin{figure}
\centering
 \includegraphics[width=.9\linewidth]{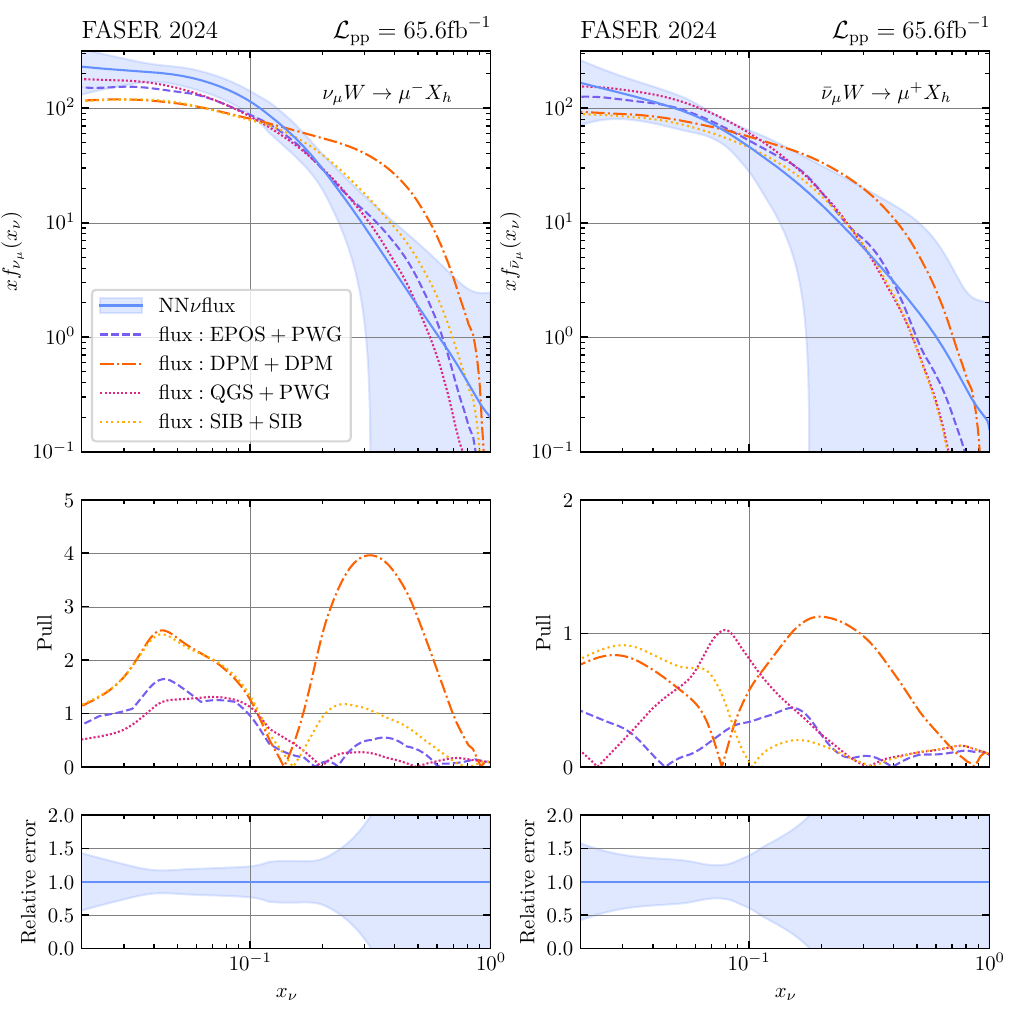}
 \caption{\label{fig:fitNN} Fit to data from the FASER experiment and 
 the estimate of the neutrino flux from different generator with 68\% CL
 uncertainties. The middle panel shows the pull between the fit and the
 prediction for the neutrino flux obtained from the four different 
 generators.}
\end{figure}

In figure \ref{fig:fitNN} a fit of the neutrino flux for the six datapoints 
measured by the FASER$\nu$ experiment \cite{FASER:2024ref} can be found, 
with the 68\% CL uncertainties.
The uncertainties coincide with the experimental uncertainties in the region
with data, but grow large in the unconstrained high-$x_\nu$ region.
Next to the fit also the simulations of the neutrino flux using the Run-3 
detector configuration \cite{Kling:2021gos,FASER:2024ykc}.
The fit shows that,  already with this small dataset, there are regions in 
which the flux generated with DPMJET can be disfavoured at a $4\sigma$ level.
The overall best agreement is found for the case in which the light hadron
production was modelled usign EPOS or QGSJET and the charmed hadrons with
POWHEG.

\section{Summary}
The comparison between the GENIE and the POWHEG-BOX implementation of neutrino
DIS has shown that there are sizeable differences and that higher-order effects
cannot be neglected.
The impact of the NLO correction, variants of parton showers and tunes for
hadronisation lead to non negligible effects.
The method to extract the flux has proven, that with just a few datapoint, this 
method exhibits a sensitivity for forward hadron production and can discriminate 
between different event generators. 

\bibliographystyle{JHEP}
\bibliography{main}

\end{document}